\documentclass[letterpaper,11pt]{JHEP3}
\usepackage{amsfonts,amsmath,latexsym,
graphicx,epsfig,cite,color}
\usepackage[english]{babel}

\newcommand{\eq}{\begin{equation}}
\newcommand{\feq}{\end{equation}}
\newcommand{\eqn}{\begin{eqnarray}}
\newcommand{\feqn}{\end{eqnarray}}

\newcommand{\ma}[1]{\mbox{$\mathcal{#1}$}}

\newcommand{\D}{{\rm d}}
\newcommand{\ti}{\tilde}

\title{Supersymmetry of the C-metric and the general Pleba\'nski-Demia\'nski solution}

\author{Dietmar Klemm$^{ab}$ and Masato Nozawa$^c$ \\
$^a$ Dipartimento di Fisica, Universit\`a di Milano, \\
\hspace*{0.15cm} Via Celoria 16, 20133 Milano, Italy. \\
$^b$ INFN, Sezione di Milano, Via Celoria 16, 20133 Milano, Italy. \\
$^c$ Theory Center, KEK, Tsukuba 305-0801, Japan.
}
\preprint{IFUM-1009-FT\\ KEK-TH-1611\\ KEK-Cosmo-114}

\abstract{We derive the necessary and sufficient conditions under which the general
Pleba\'nski-Demia\'nski (PD) solution of Einstein-Maxwell theory with a negative cosmological constant
admits Killing spinors. We consider in detail two different scaling limits of the PD metric.
The first of these limits removes the acceleration parameter, and leads to the Carter-Pleba\'nski solution.
In this case, the integrability conditions for Killing spinors were obtained by Alonso-Alberca, Meessen
and Ort\'{\i}n in hep-th/0003071, and we show that these are not only necessary, but also sufficient
for the existence of Killing spinors. This fills also a gap in hep-th/9808097, where the integrability
conditions for supersymmetry of the Kerr-Newman-AdS black hole were worked out, but the Killing
spinor was not constructed explicitely. The second scaling limit eliminates the rotation parameter, and
leads to the cosmological C-metric, which describes accelerated black holes in AdS. Also in this case,
the supersymmetry conditions are obtained, and it is shown that they follow from the ones of the general PD
solution by scaling the parameters appropriately. In all cases, we determine the three-dimensional base
space that appears in the classification scheme of hep-th/0307022, and prove that for the
1/2-supersymmetric Reissner-Nordstr\"om-AdS spacetime, this base is unique.
A Wick-rotation of our results leads to gravitational instantons that generalize the ones constructed recently
by Martelli, Passias and Sparks in arXiv:12124618 to $\text{U}(1)\times\text{U}(1)$ symmetry.
These instantons are shown to admit an integrable almost complex structure.
Finally, our work may open the possibility to systematically construct
generalizations of the PD metric that include scalar fields with a potential in matter-coupled
gauged supergravity.
}


\begin{document}

\section{Introduction}
\label{intro}

Gravitational backgrounds preserving supersymmetry in supergravity theories are central to 
the development of string/M-theory, flux compactifications and the AdS/CFT correspondence.  
Supersymmetric, or simply BPS,  solutions are characterized 
by the presence of Killing spinors $\epsilon$ which are parallel 
with respect to the supercovariant derivative operator. 
These Killing spinors define preferred G-structures which 
provide algebraic and differential constraints on the bilinears
constructed from $\epsilon$~\cite{Gauntlett:2001ur,Gauntlett:2002sc}. The classification 
program initiated in \cite{Gauntlett:2002nw} made a substantial development  
and the G-structure enables us to constrain  
the metric, fluxes and other fields to obey a simpler set of
equations~\cite{Gauntlett:2003fk,Gutowski:2004yv,Gutowski:2005id,Bellorin:2006yr,Bellorin:2007yp,Gutowski:2003rg,Gauntlett:2002fz,Gauntlett:2003wb,Caldarelli:2003pb,Bellorin:2005zc,Meessen:2010fh,Meessen:2012sr}\footnote{An alternative approach to classifying
supergravity solutions consists in expressing spinors in terms of forms and using the gauge
symmetry to transform them to a preferred representative of their orbit. This allows to directly
solve the Killing spinor equations and goes under the name of spinorial geometry,
cf.~\cite{Gran:2005wn,Gran:2005ct,Cacciatori:2007vn,Gutowski:2007ai,Grover:2008ih,Cacciatori:2008ek,Klemm:2009uw,Klemm:2010mc}
for an incomplete list of references.} (see~\cite{Maeda:2011sh} for a recent review).   

For 4-dimensional minimal $\ma N=2$ ungauged supergravity,
a complete list of BPS solutions was obtained~\cite{Tod:1983pm}.   
When the vector field constructed as bilinear of a Killing spinor is
timelike, it turned out that BPS solutions are completely specified by a complex harmonic
function on the 3-dimensional base space $\mathbb E^3$.
The BPS geometries in minimal $\ma N=2$ gauged supergravity 
were classified by Caldarelli and one
of the present authors in~\cite{Caldarelli:2003pb}, and later studied 
in~\cite{Cacciatori:2004rt,Cacciatori:2007vn}. 
It was shown that the necessary and sufficient conditions for 
supersymmetry in the timelike class reduce to the solutions of differential 
equations on a curved 3-dimensional base space. The 
striking feature of gauged supergravity is that 
these differential equations are highly nonlinear, which 
causes a main difficulty in finding solutions contrary to the ungauged case. 
The properties of supersymmetric solutions in gauged supergravities 
are therefore far from understood.

Prior to these studies, Romans analyzed asymptotically AdS static BPS solutions 
by directly solving a Killing spinor equation~\cite{Romans:1991nq}.  
Later on some rotating generalizations have been 
discussed in~\cite{Caldarelli:1998hg,AlonsoAlberca:2000cs}
by investigating the first integrability conditions for the Killing
spinor equation. It should be worthwhile to emphasize that  
integrability conditions are merely the necessary conditions for
supersymmetry \cite{vanNieuwenhuizen:1983wu}. Hence,  
in order to show rigorously that they are also sufficient, one needs to explicitly
construct the Killing spinor or to show that these bosonic
configurations fit into the classification scheme given
in~\cite{Caldarelli:2003pb}. However, the construction of Killing spinors
is difficult in the rotating case since they depend nontrivially  on 
both the radial and angular coordinates.
Thus, there remains the possibility that the integrability conditions
for supersymmetry obtained
in~\cite{Caldarelli:1998hg,AlonsoAlberca:2000cs}  may not be sufficient. 
To fill this gap in~\cite{Caldarelli:1998hg,AlonsoAlberca:2000cs}
is one of the purposes of the present paper.

The BPS backgrounds studied
in~\cite{Caldarelli:1998hg,AlonsoAlberca:2000cs} 
are contained in the class of Pleba\'nski-Demia\'nski (PD) \cite{Plebanski:1976gy},
which represents the general type D spacetime with an aligned
non-null electromagnetic field, and describes a rotating, charged and uniformly
accelerating mass. In this paper we work out the necessary and sufficient
conditions under which the PD solution is supersymmetric in minimal ${\cal N}=2$,
$D=4$ gauged supergravity. This represents our main result.
We shall also consider in detail two different scaling limits
of the PD metric. The first of these removes the acceleration parameter, and leads to
a solution discovered by Carter and Pleba\'nski \cite{Carter:1968ks,Plebanski:1975}.
In this case, the integrability conditions
for Killing spinors were obtained in \cite{AlonsoAlberca:2000cs}, and we show that
these are not only necessary, but also sufficient for supersymmetry.
The second scaling limit eliminates the rotation parameter, and leads to the
cosmological C-metric, which describes accelerating black holes in AdS.
Also in this case, the BPS conditions are obtained. For both scaling limits,
it is shown that these constraints follow from the ones of the general PD solution by
scaling the parameters appropriately. In all cases, we determine the three-dimensional
base space that appears in the classification scheme of \cite{Caldarelli:2003pb},
and prove that for the 1/2-BPS Reissner-Nordstr\"om-AdS spacetime,
this base is unique. Finally, a Wick-rotation of our results leads to gravitational
instantons with $\text{U}(1)\times\text{U}(1)$ symmetry that are supersymmetric in Euclidean
gauged supergravity. These generalize the $\text{SU}(2)\times\text{U}(1)$-symmetric ones
constructed recently in \cite{Martelli:2012sz}.

The remainder of this paper is organized as follows:
In section~\ref{sec:BPS} we review the equations obtained
in \cite{Caldarelli:2003pb,Cacciatori:2004rt} that BPS geometries in minimal ${\cal N}=2$,
$D=4$ gauged supergravity must satisfy.
Section~\ref{sec:PD} is devoted to the discussion of the above-mentioned subclasses of
the general PD solution, which arise in different scaling limits.
The BPS conditions for the general PD solution are obtained in \ref{sec:generalPD}.
In section \ref{sec:Euclidean} an analytic continuation to Euclidean signature is
considered. We conclude in section~\ref{final} with some final remarks.

\section{BPS geometries in minimal ${\cal N}=2$, $D=4$ gauged supergravity}
\label{sec:BPS}

The bosonic action of minimal gauged ${\cal N}=2$ supergravity in four
dimensions is given by
%---------------  Lagrangian   ---------------%
\begin{align}
 S=\frac{1}{16\pi G}\int \D ^4 x\sqrt{-g}
\left(R-2 \Lambda -{\cal F}_{\mu\nu}{\cal F}^{\mu\nu}\right)\,, \label{action}
\end{align}
where $\Lambda =-3\ell^{-2} (<0)$ and ${\cal F}=\D{\cal A}$. 
A bosonic configuration is said to be supersymmetric if 
it admits a Killing spinor $\epsilon $ satisfying 
%-------------- Killing spinor   --------------%
\begin{align}
\hat\nabla_{\mu}\epsilon \equiv \left(\nabla_\mu +\frac{ i}4 \ma F_{\nu\rho }
\Gamma^{\nu\rho }\Gamma_\mu +\frac{1}{2\ell }\Gamma_\mu 
-\frac{ i}{\ell}\ma A_\mu \right)\epsilon =0\,. \label{Kill-spin}
\end{align}
The existence of a Killing spinor imposes strong restrictions on the geometry
and the Maxwell field. The key role is played by bilinears 
constructed from a Killing spinor, 
\begin{align}
f:=\bar\epsilon \epsilon\,,\qquad 
g:=i \bar\epsilon\Gamma_5 \epsilon \,, \qquad 
V_\mu :=i\bar \epsilon \Gamma_\mu \epsilon \,, 
\qquad A_\mu: =i \bar \epsilon \Gamma_5 \Gamma_\mu \epsilon \,, \qquad 
\Phi_{\mu\nu } :=i \bar \epsilon \Gamma_{\mu\nu }\epsilon \,,
\label{bilinears}
\end{align}  
where $\bar \epsilon =i\Gamma^0 \epsilon ^\dagger $ and
$\Gamma_5=i\Gamma_{0123}$. It turns out that the vector $V_\mu $ is a causal Killing
field, so that the general BPS solutions fall into two categories, namely a timelike
and a null family.
The general timelike supersymmetric solution of
this theory was obtained in \cite{Caldarelli:2003pb}, and reads
(cf.~also \cite{Cacciatori:2004rt})
%--------------  SUSY metric  ---------------%
\begin{eqnarray}
\D s^2 &=& -\frac4{\ell^2F\bar F}(\D t + \omega_i\D x^i)^2 + \frac{\ell^2F\bar F}4\left[\D z^2 + e^{2\phi}(\D x^2
              + \D y^2)\right]\,, \label{metric-BPS} \\
{\cal F} &=& \frac{\ell^2}4 F\bar F\left[V\wedge \D f + \star\left(V\wedge\left(\D g + \frac1{\ell}\D z\right)
                      \right)\right]\,, \label{fluxes-BPS}
\end{eqnarray}
where $i=1,2$; $x^1=x$, $x^2=y$, and we defined $\ell F=2i/(f-ig)$. 
The timelike class of solutions preserves at least one quarter of the supersymmetry. 
In the canonical form \eqref{metric-BPS}, we have $V^\mu =(\partial_t)^\mu$
and $A_\mu =\nabla_\mu z$.
The real function
$\phi$ and the complex function $F$, that depend only on $x,y,z$, are
determined by the system
%--------------  SUSY system  ---------------%
\begin{eqnarray}
\Delta F + e^{2\phi}[F^3 + 3FF' + F''] &=& 0\,, \label{Delta-F} \\
\Delta\phi + \frac12 e^{2\phi}[F' + \bar F' + F^2 + \bar F^2 - F\bar F] &=& 0\,, \label{Delta-phi} \\
\phi' - \text{Re}F &=& 0\,, \label{phi'}
\end{eqnarray}
where $\Delta=\partial^2_x+\partial^2_y$, and a prime denotes
differentiation with respect to $z$.
Eq. (\ref{Delta-F}) comes from the Maxwell equations and the Bianchi
identity, (\ref{Delta-phi}) from the integrability condition of a
Killing spinor equation and (\ref{phi'}) from the 
differential conditions for bilinears. 
Finally, the one-form $\omega$ is obtained from\footnote{We have chosen an axial gauge in which
$\omega_z=0$. The integrability of (\ref{omega}) is guaranteed by
(\ref{Delta-F}).  
}
%--------------  omega  ---------------%
\begin{eqnarray}
\partial_z\omega_i &=& \frac{\ell^4}8(F\bar F)^2\epsilon_{ij}(f\partial_j g - g\partial_j f)\,, \nonumber \\
\partial_i\omega_j - \partial_j\omega_i &=& \frac{\ell^4}8(F\bar F)^2 e^{2\phi}\epsilon_{ij}\left(f\partial_z g
- g\partial_z f + \frac{2f}{\ell}\right)\,, \label{omega}
\end{eqnarray}
with $\epsilon_{12}=1$. Decomposing $F$ into its real and imaginary part, $F=A+iB$, we see that
the real part of eqn.~\eqref{Delta-F} follows from \eqref{Delta-phi} and \eqref{phi'}, so that the remaining
system is
\begin{eqnarray}
\Delta B + e^{2\phi}[3\phi'^2B - B^3 + 3\phi'B' + 3B\phi'' + B''] &=& 0\,, \label{Delta-B} \\
\Delta\phi + \frac12 e^{2\phi}[2\phi'' + {\phi'}^2 - 3B^2] &=& 0\,, \label{Delta-phi-2}
\end{eqnarray}
together with $A=\phi'$.

A notable feature of supersymmetric solutions in gauged supergravity
is that the system obeys the nonlinear set of equations (\ref{Delta-F})--(\ref{phi'}). 
This is in sharp contrast with the ungauged
theory \cite{Tod:1983pm},
where the BPS solutions in the timelike class are specified by 
harmonics on the 3-dimensional base space $\mathbb E^3$. 
This is a major obstacle one encounters in attempting to find 
supersymmetric solutions in the gauged case. 

Another difficulty in gauged supergravity is that the Killing vector 
$V^\mu =i\bar \epsilon \gamma^\mu \epsilon $ is not associated with the 
natural time translation of AdS space. This fact can be seen by 
solving the system  (\ref{Delta-F})--(\ref{phi'}) for $\ma F_{\mu\nu }=0$. 
Setting $\ma F_{\mu\nu }=0$ leads to $f={\rm constant}$, thereby 
we can set $f=0$ or $1$ without loss of generality. The former case 
gives AdS in Poincar\'e coordinates, which do not cover
AdS globally, whereas the latter case yields~\cite{Caldarelli:2003pb}
\begin{align}
 \D s^2
 =\frac{\D z^2}{1+z^2/\ell^2 }
+\left(1+\frac{z^2}{\ell^2}\right) 
\left[-\left(\D t+\ell\sinh^2\frac{\vartheta }{2}
\D \varphi \right)^2
+\frac{\ell^2 }{4} (\D \vartheta^2 +\sinh^2 \vartheta \D \varphi^2) 
\right]\,.\label{AdS_SUSYSL2R}
\end{align}
The level set $z={\rm constant}$ represents  AdS$_3$
written in an ${\rm SL}(2,\mathbb R)$ invariant form.
The embedding into $\mathbb E^{3,2}$ (with signature
($-1,1,1,1,-1$)) is given by
\begin{align}
 X^0& =\ell \cos(t/\ell)\cosh(\vartheta/2) \sqrt{1+ z^2/\ell^2} \,,
 \nonumber \\
X^1 &= \ell \cos (\varphi-t/\ell)\sinh(\vartheta/2)\sqrt{1+ z^2/\ell^2} 
\,,
\nonumber \\
X^2 &= \ell \sin (\varphi-t/\ell)\sinh(\vartheta/2)\sqrt{1+ z^2/\ell^2}\,,
 \\ 
X^3 &= z \,, \nonumber \\
X^4 &=\ell \sin(t/\ell) \cosh(\vartheta/2) \sqrt{1+ z^2/\ell^2}
 \,. \nonumber 
\end{align}
Hence the coordinate system ($t, z, \vartheta, \varphi$) 
covers AdS globally and the metric can be brought into the standard form 
\begin{align}
 \D s^2 = -\left(1+\frac{r^2}{\ell^2}\right)\D {\ti t}^2 
+ \frac{\D r^2}{1+r^2/\ell^2}
+r^2 (\D \theta ^2+\sin^2\theta \D {\phi } ^2)
\end{align}
by the coordinate transformation
\begin{align}
t =\ti t\,, \qquad \varphi=\phi+\frac{\ti t}{\ell } \,, \qquad 
z=r \cos\theta \,, \qquad \sinh\frac{\vartheta }{2} 
=\frac{r\sin\theta }{\ell\sqrt{1+(r/\ell)^2 \cos^2\theta }} \,. 
\label{AdScharts_static-susy}
\end{align}
Inspection of (\ref{AdScharts_static-susy}) 
leads to 
\begin{align}
\frac{\partial }{\partial \ti t}=\frac{1}{\ell } \left( 
 X^0\frac{\partial }{\partial X^4 }-X^4\frac{\partial }{\partial X^0}
\right)
=\frac{\partial }{\partial t}+\frac{1}{\ell}\frac{\partial }{\partial
 \varphi} \,.  
\label{AdS_globa_SUSY_tr}
\end{align}
This means that an observer following orbits of the Killing vector associated with the Killing spinor 
is rotating by the constant angular velocity $\ell^{-1}$ with respect to the static observer at AdS infinity.
This is in accordance with the fact that the Bogomol'nyi bound for
($\ma N=2$) gauged supergravity involves the angular momentum~\cite{Gibbons:1983aq,Kostelecky:1995ei}, 
\begin{align}
 M\ge \frac{1}{\ell }|J| +\mathsf Q \,, 
\end{align}
where $M $ and $J $ are the Abott-Deser mass and angular
momentum respectively \cite{Abbott:1981ff}, and 
$\mathsf Q$ denotes the electric charge\footnote{Note that the magnetic charge does not enter
the ${\rm osp}(4|2)$ superalgebra since it breaks ${\rm SO}(3,2)$ covariance~\cite{Dibitetto:2010sp}. The issue of BPS bounds in minimal ${\cal N}=2$, $D=4$ gauged supergravity has recently
been studied and clarified in \cite{Hristov:2011ye}.}. 
The same remark applies also to the BPS solutions in  five dimensions, where for instance
AdS$_5$ is represented by a fibration over the Bergmann manifold~\cite{Gauntlett:2003fk}. 

The above instance illustrates that well-known static BPS solutions   
may be expressed in a rotating frame in the formulation of \cite{Caldarelli:2003pb}. 
This raises an additional obstacle to obtain BPS solutions.  
Bearing these remarks in mind, we shall show below how to derive 
supersymmetric solutions.

\section{The Pleba\'nski-Demia\'nski solution}
\label{sec:PD}
The complete family of type-D spacetimes with a non-null electromagnetic field, whose two principal
null congruences are aligned with the two repeated principal null congruences of the Weyl tensor, was
given by Pleba\'nski and Demia\'nski \cite{Plebanski:1976gy}\footnote{For a more recent review
cf.~\cite{Griffiths:2005qp}.}. It solves the field equations of
Einstein-Maxwell-(A)dS gravity and describes a rotating, charged and
uniformly accelerating mass.
The metric and field strength read respectively
%-----------------  PD metric  ------------------%
\begin{eqnarray}
\D s^2 = &&
\frac1{(1-pq)^2}\left\{-\frac{Q(q)}{p^2+q^2}(\D\tau -p^2\D\sigma)^2 
+ \frac{p^2+q^2}{Q(q)}\D q^2
\right.\nonumber \\ &&
+ \left.\frac{p^2+q^2}{P(p)}\D p^2 
+ \frac{P(p)}{p^2+q^2}(\D\tau + q^2\D\sigma)^2\right\}\,, 
\label{metr-PD}
\end{eqnarray}
\eq
{\cal F} = \frac{\mathsf{Q}(p^2-q^2)+2\mathsf{P}pq}{(p^2+q^2)^2}\D q\wedge (\D\tau - p^2\D\sigma)
+ \frac{\mathsf{P}(p^2-q^2)-2\mathsf{Q}pq}{(p^2+q^2)^2}\D p\wedge (\D\tau + q^2\D\sigma)\,, \label{F-PD}
\feq
where the structure functions are given by
\begin{eqnarray}
P(p) &=& (-\Lambda/6 - \mathsf{P}^2 + \alpha) + 2np - \varepsilon p^2 + 2mp^3 + (-\Lambda/6 -
                  \mathsf{Q}^2 - \alpha)p^4\,, \nonumber \\
Q(q) &=& (-\Lambda/6 + \mathsf{Q}^2 + \alpha) - 2mq + \varepsilon q^2 - 2nq^3 + (-\Lambda/6 +
                  \mathsf{P}^2 - \alpha)q^4\,.
\end{eqnarray}
%Our mathsf Q (which is also the one of Tomas ) is -e of PD; our mathsf P is g of PD. Siehe Zettel 1), 2) vom 27.7.12.
Here, $\alpha,\varepsilon,m,n,\mathsf{P},\mathsf{Q}$ are arbitrary parameters, with $\mathsf{P}$ and
$\mathsf{Q}$ representing the magnetic and electric charges
respectively. 
Eqn.~\eqref{metr-PD} together with~\eqref{F-PD} solve the equations of motion following from~\eqref{action}.

The main purpose of this section is to determine the condition under
which the general PD solution preserves supersymmetry.

\subsection{First scaling limit: The Carter-Pleba\'{n}ski metric}

A subclass of solutions can be obtained by scaling the coordinates according to
\eq
p \to l^{-1}p\,, \qquad q \to l^{-1}q\,, \qquad \tau \to l\tau\,, \qquad \sigma \to l^3\sigma\,, \label{scaling1-coord}
\feq
and simultaneously adjusting the constants
\eq
\mathsf{P} \to l^{-2}\mathsf{P}\,, \quad \mathsf{Q} \to l^{-2}\mathsf{Q}\,, \quad m \to l^{-3}m\,, \quad
n \to l^{-3}n\,, \quad \varepsilon \to l^{-2}\varepsilon\,, \quad \alpha \to l^{-4}\alpha + \Lambda/6\,,
\label{scaling1-param}
\feq
and taking the limit $l\to\infty$. This removes the acceleration parameter\footnote{The acceleration
parameter is essentially given by $l^{-2}$, as can be seen by comparing \eqref{scaling1-coord} and
\eqref{scaling1-param} with eqns.~(3) and (4) of \cite{Griffiths:2005qp}.} and leads
to~\cite{Plebanski:1976gy}
\eq
\D s^2 = -\frac{Q(q)}{p^2+q^2}(\D\tau - p^2\D\sigma)^2 + \frac{p^2+q^2}{Q(q)}\D q^2
                + \frac{p^2+q^2}{P(p)}\D p^2 + \frac{P(p)}{p^2+q^2}(\D\tau + q^2\D\sigma)^2\,, 
                      \label{metr-PD-scaled}
\feq
\begin{eqnarray}
P(p) &=& \alpha - \mathsf{P}^2 + 2np - \varepsilon p^2  + (-\Lambda/3)p^4\,, \nonumber \\
Q(q) &=& \alpha + \mathsf{Q}^2 - 2mq + \varepsilon q^2 + (-\Lambda/3)q^4\,. \label{struc-func-scaled}
\end{eqnarray}
The electromagnetic field is still given by \eqref{F-PD}. In what follows, we shall refer to
\eqref{metr-PD-scaled} as the Carter-Pleba\'nski solution, since it was derived and studied already by
Carter \cite{Carter:1968ks} and later by Pleba\'nski \cite{Plebanski:1975}.
Notice that one can take a different scaling limit (after the inversion $q\to-1/q$), leading to the
cosmological C-metric, which will be considered in the next subsection.

The first integrability conditions for \eqref{metr-PD-scaled} to admit Killing spinors as a solution to minimal
gauged ${\cal N}=2$ supergravity were analyzed in \cite{AlonsoAlberca:2000cs}. There it was found that
they are equivalent to
%-------------  BPS bound  -----------------%
\eq
\frac1{\ell}[m\mathsf{P} + n\mathsf{Q}] = 0\,, 
\qquad {\cal B}_+{\cal B}_- = 0\,, \label{integr-cond-PD}
\feq
where
\eq
{\cal B}_{\pm} \equiv m^2 + n^2 - (\varepsilon\pm 2\alpha^{1/2}/\ell)(\mathsf{P}^2 + \mathsf{Q}^2)\,.
\feq
As is well-known~\cite{vanNieuwenhuizen:1983wu}, in general the integrability conditions are
necessary but not sufficient for a solution to admit Killing spinors. (An explicit counterexample
was given in~\cite{Cacciatori:2004rt}). However, for the case of the
Carter-Pleba\'nski solution~\eqref{metr-PD-scaled}, 
we will show below that the conditions~\eqref{integr-cond-PD} are not only necessary but also
sufficient. 

To this aim, we show
how to obtain the (supersymmetric) Carter-Pleba\'nski solution from the equations \eqref{Delta-B},
\eqref{Delta-phi-2}.  
It turns out that the correct way to do this is to define new
coordinates $q,p$ by
%---------------- (q, p)   --------------------%
\eq
x= \ell [\alpha(q) + \beta(p)]\,, \qquad 
z ={\ell} \gamma(q)\delta(p)\,,
\feq
where $\gamma(q)=\gamma_0+\gamma_1 q$, $\delta(p)=\delta_0+\delta_1 p$, and $\gamma_0$,
$\gamma_1$, $\delta_0$, $\delta_1$ are real constants. By rescaling $q$ and $p$ one can always
set $\gamma_1=\delta_1={\ell^{-1}}$. The function $\phi$ is assumed
to be separable,
%-------------- exp(2\phi)   ----------------%
\eq
e^{2\phi} = \rho(q)\psi(p)\,, \label{sep-ansatz-phi}
\feq
where $\rho$ and $\psi$ are both fourth-order polynomials,
%------------------ (\rho,\psi)   --------------------%
\eq
\rho(q) = \sum_{n=0}^4\rho_n q^n\,, \qquad \psi(p) = \sum_{n=0}^4\psi_n p^n\,.
\feq
Finally, $\alpha$, $\beta$ are determined by requiring that there be no mixed terms $\sim\D p\D q$ in
the base space metric
%------------------ 3D base   --------------------%
\eq
\D s_3^2 = \D z^2 + e^{2\phi}(\D x^2 + \D y^2)\,.
\feq
This yields
%------------------ (alpha, beta)   --------------------%
\eq
\alpha'(q) = -\frac{\gamma}{{\ell}\rho}\,, \qquad \beta'(p) = \frac{\delta}{{\ell}\psi}\,,
\feq
where a prime denotes differentiation w.r.t.~the corresponding
argument. Then, eq. \eqref{Delta-phi-2}
allows to compute $B$, with the result
%------------------  B^2  --------------------%
\eq
3B^2 = \left(\gamma^2\psi+\delta^2\rho\right)^{-2}\left[(\rho'' + \psi'')(\gamma^2\psi + \delta^2\rho)
              -\frac34(\gamma\psi' + \delta\rho')^2\right]\,. \label{3B^2}
\feq
Obviously, the final metric will be rather complicated unless the expression on the rhs of \eqref{3B^2}
is a perfect square. It can be checked that this is the case if and only if the following relations for
the coefficients hold:
%------------------  Parameters  --------------------%
\eq
\psi_2 = -\rho_2\,, \qquad \psi_3 = \rho_3 = 0\,, \qquad \psi_4 = \rho_4\,, \qquad \delta_0 =
-\frac{\gamma_0\psi_1}{\rho_1}\,,
\feq
\eq
\rho_0 = \frac{{\ell}\gamma_0\rho_1}2 + \frac{\lambda^2}{4\rho_4}\,, \qquad \psi_0 = -\frac{{\ell}\gamma_0\psi_1^2}
{2\rho_1} + \frac{\lambda^2}{4\rho_4}\,,
\feq
where we defined
\eq
\lambda = \frac{\rho_1}{2\gamma_0{\ell}} - \rho_2\,.
\feq
In this case,
\begin{displaymath}
(\rho'' + \psi'')(\gamma^2\psi + \delta^2\rho) - \frac34(\gamma\psi' + \delta\rho')^2 =
3\left[a_1p + a_2p^2 + q(b_0 + b_2p^2) + q^2(c_0 + c_1p)\right]^2\,,
\end{displaymath}
with
\begin{displaymath}
a_1 = -\frac{\psi_1\gamma_0\lambda}{\rho_1}\,, \qquad a_2 = c_0 = \frac{\lambda}{{\ell}}\,, \qquad b_0 = \gamma_0
\lambda\,, \qquad b_2 = 2\gamma_0\rho_4\,, \qquad c_1 = \frac{2\gamma_0\psi_1\rho_4}{\rho_1}\,,
\end{displaymath}
and thus
%------------------  B  --------------------%
\eq
B = \frac1{\gamma^2\psi+\delta^2\rho}\left[-\frac{\psi_1\gamma_0\lambda}{\rho_1}p + \frac{\lambda}{{\ell}} p^2
+ q\gamma_0(\lambda + 2\rho_4 p^2) + q^2\left(\frac{\lambda}{{\ell}} + \frac{2\gamma_0\psi_1\rho_4}{\rho_1}
p\right)\right]\,. \label{B}
\feq
Remarkably, one finds that then eqn.~\eqref{Delta-B} is automatically satisfied. It would be very
interesting to understand if there is a deeper reason for this.

Given $\phi$ and
$B$, the function $F$ can be computed from $F=\partial_z\phi+iB$. Finally, the one-form $\omega$
is obtained by integrating \eqref{omega}, with the result $\omega=\omega_y\D y$, where
\eq
\omega_y = -\frac{{\ell^3}\rho_1}{2\gamma_0}\left[\frac{\rho\left(p^2+\frac{\lambda}{2\rho_4}
\right) + \psi\left(q^2-\frac{\lambda}{2\rho_4}\right)}{\rho\left(p^2+\frac{\lambda}{2\rho_4}\right)^2
- \psi\left(q^2-\frac{\lambda}{2\rho_4}\right)^2} +c \sqrt{\frac{2\gamma_0\rho_4}{{\ell}\rho_1}}\right]\,.
\label{omega_y}
\feq
Note that the integration constant in \eqref{omega_y} was chosen for
later convenience and the dimensionless constant $c$ was inserted 
to take limits in the subsequent sections. If we introduce
new coordinates $\tau,\sigma$ according to
\begin{align}
\left(\begin{array}{c} t \\ y \end{array}\right) = \left(
\begin{array}{cc} c & \sqrt{\frac{{\ell}\rho_1}{2\gamma_0\rho_4}}
+ \frac{\lambda c}{2\rho_4} \\ \frac1{\rho_4\ell^2}\sqrt{\frac{2\gamma_0\rho_4}{{\ell}\rho_1}} & \frac{\lambda}
{2\rho_4^2\ell^2}\sqrt{\frac{2\gamma_0\rho_4}{{\ell}\rho_1}} 
\end{array}\right)\left(\begin{array}{c} \tau \\
\sigma \end{array}\right)\,,
\end{align}
the metric \eqref{metric-BPS} becomes
\eq
\D s^2 = -\frac{Q(q)}{p^2+q^2}(\D\tau - p^2\D\sigma)^2 + \frac{p^2+q^2}{Q(q)}\D q^2 + \frac{p^2+q^2}{P(p)}
\D p^2 + \frac{P(p)}{p^2+q^2}(\D\tau + q^2\D\sigma)^2\,, \label{metric-PD-susy}
\feq
with the structure functions
\eq
Q = \frac{\rho}{\ell^2\rho_4}\,, \qquad P = \frac{\psi}{\ell^2\rho_4}\,. \label{struc-func-susy}
\feq
The fluxes can be computed from \eqref{fluxes-BPS}, which yields
\eq
{\cal F}_{01} = \frac{\mathsf{Q}(q^2-p^2) - 2\mathsf{P}pq}{(p^2 + q^2)^2}\,, \qquad
{\cal F}_{23} = -\frac{\mathsf{P}(q^2-p^2) + 2\mathsf{Q}pq}{(p^2 + q^2)^2}\,, \label{fluxes-susy}
\feq
where the electric and magnetic charges are given respectively by
\eq
\mathsf{Q} = -\frac{\gamma_0}{\ell}\sqrt{\frac{\rho_1 {\ell}}{2\gamma_0\rho_4}}\,, \qquad
\mathsf{P} = -\frac{\gamma_0\psi_1}{\ell\rho_1}\sqrt{\frac{\rho_1 {\ell}}{2\gamma_0\rho_4}}\,, \label{charges-susy}
\feq
and we have chosen the tetrad
\[
e^0 = \left(\frac{Q}{p^2+q^2}\right)^{1/2}(\D\tau - p^2\D\sigma)\,, \qquad e^1 = \left(\frac{p^2+q^2}{Q}
\right)^{1/2}\D q\,,
\]
\[
e^2 = \left(\frac{p^2+q^2}{P}\right)^{1/2}\D p\,, \qquad
e^3 = \left(\frac{P}{p^2+q^2}\right)^{1/2}(\D\tau + q^2\D\sigma)\,.
\]
Eqns.~\eqref{metric-PD-susy} and \eqref{fluxes-susy} coincide precisely with \eqref{metr-PD-scaled}
and \eqref{F-PD}, apart from the obvious fact that the structure functions $Q,P$ in \eqref{struc-func-susy}
are more restricted than \eqref{struc-func-scaled} due to supersymmetry. Comparing
\eqref{struc-func-susy} with \eqref{struc-func-scaled}, we obtain for the parameters
$\varepsilon,\alpha,m,n$
\eq
\varepsilon = \frac{\rho_2}{\ell^2\rho_4}\,, \qquad \alpha = \frac1{4\ell^2\rho_4^2}\left(\frac{\rho_1}
{2\gamma_0 {\ell}} - \rho_2\right)^2\,, \qquad m = -\frac{\rho_1}{2\ell^2\rho_4}\,, \qquad n = \frac{\psi_1}
{2\ell^2\rho_4}\,. \label{constants-susy}
\feq
The BPS solution is therefore completely specified by the four constants $\gamma_0$, $\rho_1/(\ell^2\rho_4)$,
$\rho_2/(\ell^2\rho_4)$ and $\psi_1/(\ell^2\rho_4)$. One easily verifies that the charges
\eqref{charges-susy}, together with the parameters \eqref{constants-susy}, do indeed satisfy the
conditions \eqref{integr-cond-PD} found in \cite{AlonsoAlberca:2000cs}, where
${\cal B}_+=0$ if $\rho_1/(2\gamma_0\ell)-\rho_2>0$ and ${\cal B}_-=0$ if $\rho_1/(2\gamma_0\ell)-\rho_2<0$.
We have thus confirmed that the supersymmetric  Carter-Pleba\'nski spacetime must obey \eqref{integr-cond-PD},
but the converse is also true: The two constraints \eqref{integr-cond-PD} leave four free constants
out of the six $\mathsf{Q},\mathsf{P},\varepsilon,\alpha,m,n$. Since the BPS solution that we obtained here
contains also four parameters, the integrability conditions \eqref{integr-cond-PD} are not only necessary,
but also sufficient for the existence of a Killing spinor.

\subsubsection{Kerr-Newman-AdS}

The Kerr-Newman-AdS (KNAdS) spacetime can be obtained from the Carter-Pleba\'nski solution
\eqref{metr-PD-scaled} by setting
\begin{align}
 &
p = a\cos\theta\,, \qquad q = r\,, \qquad 
\tau = \ti t - \frac{a\phi}{\Xi}\,, \qquad 
\sigma = -\frac{\phi}{a\Xi}\,,
\nonumber \\
&n = 0\,, \qquad \varepsilon = 1 + \frac{a^2}{\ell^2}\,, \qquad
\alpha = a^2 + \mathsf{P}^2\,,
\end{align}
which yields the KNAdS metric in Boyer-Lindquist coordinates,
\begin{align}
\D s^2=-\frac{\Delta }{\Sigma^2}\left(\D \ti t
-\frac{a\sin ^2 \theta}{\Xi}
\D \phi \right)^2+\frac{\Sigma^2}{\Delta }\D r^2+\frac{\Sigma^2}{\Delta
 _\theta }\D \theta ^2+\frac{\Delta _\theta \sin ^2 \theta}
{\Sigma^2}\left(a\D \ti t -\frac{r^2+a^2}{\Xi}\D\phi\right)^2,
\end{align}
where
\begin{align*}
\Sigma^2 &=r^2+a^2 \cos ^2 \theta\,, &\Xi &=1-a^2\ell ^{-2}\,, \nonumber \\
\Delta &=(r^2+a^2)(1+\ell ^{-2}r^2) - 2mr + \mathsf{Q}^2 + \mathsf{P}^2\,,
&\Delta _\theta &=1-a^2\ell^{-2}\cos ^2\theta\,.
\end{align*}
In this subcase, the necessary and sufficient conditions for supersymmetry \eqref{integr-cond-PD} boil
down to
\[
m\mathsf{P} = 0\,, \qquad {\cal B}_+{\cal B}_- = 0\,,
\]
with
\[
{\cal B}_{\pm} = m^2 - \left[1 + \frac{a^2}{\ell^2} \pm\frac2{\ell}(a^2 + \mathsf{P}^2)^{1/2}\right]
(\mathsf{P}^2 + \mathsf{Q}^2)\,.
\]
For $m=0$ we have thus the Dirac-type condition
\begin{align}
\left(1 - \frac{a^2}{\ell^2}\right)^2 = \frac4{\ell^2}\mathsf{P}^2\,,
\label{BPScond_KNAdS1}
\end{align}
which is precisely eqn.~(98) of \cite{Caldarelli:1998hg}, while for $\mathsf{P}=0$ we get
\begin{align}
m^2 = \left(1\pm\frac a{\ell}\right)^2\mathsf{Q}^2\,,
\label{BPScond_KNAdS2}
\end{align}
i.e., eqn.~(93) of \cite{Caldarelli:1998hg}. This fills a gap in \cite{Caldarelli:1998hg}, where only
the first integrability conditions for Killing spinors of the KNAdS black hole were considered.
In a similar way one can easily verify that the integrability conditions for Killing spinors of the rotating
cylindrical or hyperbolic black holes (which also arise as subcases of the general metric
\eqref{metr-PD-scaled}) given in \cite{Caldarelli:1998hg} are sufficient as well.

It is obvious that (\ref{BPScond_KNAdS1}) implies no event horizon, whereas
eqn. (\ref{BPScond_KNAdS2}) leads to 
\begin{align}
\Delta =\left(\frac{r^2}{\ell}\mp a \right)^2+
\left[\mathsf Q -\left(1\pm \frac{a}{\ell}\right)r\right]^2\,.
\end{align}
Therefore, the supersymmetric KNAdS metric 
describes a naked singularity unless 
$\mathsf Q= \sqrt{|a\ell |}(1\pm a/\ell)$, which 
provides a degenerate horizon.

\subsubsection{Reissner-Nordstr\"om-AdS}

It is also enlightening to examine the 
Reissner-Nordstr\"om-AdS (RNAdS) limit. In~\cite{Romans:1991nq},
it was shown by direct integration of the Killing spinor equations
that the RNAdS solution has a supersymmetric limit. 
The metric of the supersymmetric RNAdS space-time is given
by~\cite{Romans:1991nq}
%----------------  RNAdS  --------------------%  
\eq
\D s^2 = -U^2(r)\D \ti t^2 + U^{-2}(r)\D r^2 + r^2(\D\theta^2 + \sin^2\!\theta \D\phi^2)\,, \label{metric-RNAdS-susy}
\feq
where
\eq
U(r) = \left[\left(1 - \frac Mr\right)^2 + \frac{r^2}{\ell^2}\right]^{1/2}\,.
\feq
It admits the Killing spinor \cite{Romans:1991nq}
\eq
\epsilon = \exp\left(\frac{i\ti t}{2\ell}\right)\left(\cos\frac{\theta}2 + i\Gamma_{012}\sin\frac{\theta}2\right)
\left(\cos\frac{\phi}2 + \Gamma_{23}\sin\frac{\phi}2\right)\tilde\epsilon(r)\,,
\feq
with
\eq
\tilde\epsilon(r) = \left(\sqrt{U(r) + r/\ell} + i\Gamma_0\sqrt{U(r) - r/\ell}\right)P(-\Gamma_1)\epsilon_0\,.
\feq
Here, $\epsilon_0$ is a constant Dirac spinor, and $P(-\Gamma_1)=(1-\Gamma_1)/2$ is a projector
that reduces the solution space from four to two complex dimensions, i.e., \eqref{metric-RNAdS-susy}
preserves half of the supersymmetries. We have thus two linearly independent Killing spinors, each
of which leading to a set of bilinears \eqref{bilinears} that determines a fibration \eqref{metric-BPS} over
a three-dimensional base space. In order to determine the most general three-dimensional base, let us
compute these bilinears.

Taking into account that  ${\Gamma^0}^{\dagger}=-\Gamma^0$, ${\Gamma^i}^{\dagger}=\Gamma^i$
$(i=1,2,3)$, $\bar\epsilon=i\epsilon^{\dagger}\Gamma^0$, we obtain
\begin{align}
f =- 2\left(1 - \frac Mr\right)\zeta^{\dagger}\zeta\,,
\end{align}
where we defined $\zeta\equiv P(-\Gamma_1)\epsilon_0$, and 
\eq
g= \frac{2i r}{\ell}\zeta^{\dagger}\left(\Gamma_{23}\cos\theta -
i\Gamma_{03}\sin\theta\cos\phi + i\Gamma_{02}\sin\theta\sin\phi\right)\zeta\,. \label{g}
\feq
In order to simplify \eqref{g} further, decompose
\[
\zeta = \zeta_+ + \zeta_-\,, \qquad \zeta_{\pm} = \frac12(1 \pm i\Gamma_{23})\zeta \equiv
P(\pm i\Gamma_{23})\zeta\,.
\]
Note that $P(\pm i\Gamma_{23})$ are projectors that commute with $P(-\Gamma_1)$. The spinors
$\zeta_{\pm}$ have thus each one independent complex component. This yields
\eq
g=\frac{2r}{\ell}\left[(\zeta_+^{\dagger}\zeta_+ - \zeta_-^{\dagger}\zeta_-)
\cos\theta + i(\zeta_+^{\dagger}\Gamma_{02}\zeta_- \,e^{i\phi} - \zeta_-^{\dagger}\Gamma_{02}\zeta_+\,
e^{-i\phi})\sin\theta\right]\,.
\feq
For the timelike Killing vector $V$ and the closed one-form $A$ one gets
\begin{align}
V^0&=2U \zeta^\dagger \zeta \,, \qquad
 V^1 =0\,, \qquad  
V^2 =\frac{2r}{\ell}(e^{-i\phi}\zeta_-^\dagger
 \Gamma_{02}\zeta_++e^{i\phi}\zeta_+^\dagger \Gamma_{02}\zeta_-
 )\,,\nonumber \\
V^3 &=-\frac{2r}{\ell} \left[\sin\theta  (\zeta_+^\dagger \zeta_+ -\zeta_-^\dagger
 \zeta_-)  + i\cos\theta  (e^{-i\phi}\zeta_-^\dagger \Gamma_{02}\zeta_+
-e^{i\phi} \zeta_+^\dagger \Gamma_{02} \zeta_-
 )\right]\,,\nonumber \\
A^0&=0\,, \qquad 
A^1=2 U \left[-\cos\theta (\zeta _+^\dagger
 \zeta_+-\zeta_-^\dagger \zeta_- )+i\sin\theta
(e^{-i\phi}\zeta_-^\dagger \Gamma_{02}\zeta_+ -e^{i\phi}\zeta_+^\dagger
\Gamma_{02} \zeta_-)  \right] \,, \nonumber \\
A^2 &= 2\left(1-\frac{M}{r}\right) \left[\sin\theta (\zeta _+^\dagger
 \zeta_+-\zeta_-^\dagger \zeta_- )+i\cos\theta (e^{-i\phi}
 \zeta_-^\dagger \Gamma_{02 }\zeta_+-e^{i\phi}
 \zeta_+^\dagger \Gamma_{02 }\zeta_-) \right]
\,, \nonumber \\
 A^3& =2 \left(1-\frac{M}{r}\right) (e^{-i\phi}\zeta_-^\dagger
 \Gamma_{02 }\zeta_++e^{i\phi}
 \zeta_+^\dagger \Gamma_{02 }\zeta_-) \,, 
\label{RNAdS_bilinears}
\end{align}
where we have defined the tetrad frame as 
\begin{align}
 e^0=U \D\ti t \,, \qquad e^1=U^{-1} \D r \,, \qquad e^2 =r \D \theta \,,
 \qquad e^3 =r \sin\theta \D \phi \,.  
\end{align}
Let us normalize    
$\zeta ^\dagger \zeta =1/2$ and  define the constants
\begin{align}
c_1=2(i \zeta_-^\dagger \Gamma_{02}\zeta_++{\rm c.c}) \,, \qquad 
c_2=-2 (\zeta_-^\dagger \Gamma_{02}\zeta _++{\rm c.c})\,, \qquad 
c_3=2(\zeta_+^\dagger \zeta_+-\zeta_-^\dagger \zeta_-) \,,
\end{align}
It then follows that 
$\sum_ic_i^2=1$ and 
\begin{align}
f&=-\left(1-\frac{M}{r}\right) \,, \qquad \qquad \quad ~
g=\frac{r}{\ell } [\sin\theta (-c_1\cos\phi +c_2 \sin \phi )+c_3
 \cos\theta ] \,, 
\nonumber \\
V&=\partial_{\ti t} -\frac{1}{\ell}\sum_i c_i \xi_i \,, \qquad 
A=\D \left[(r-M)\{(c_1\cos\phi -c_2 \sin\phi )\sin\theta -c_3 \cos\theta
 \}\right]\,, \label{RNAdS_bilinears2}
\end{align}
where the $\xi_i$ denote ${\rm SO}(3)$ Killing vectors,
\begin{align}
\xi_1 =\sin\phi \partial_\theta +\cot\theta 
\cos\phi \partial_\phi \,, \qquad 
\xi_2 =\cos\phi \partial_\theta -\cot\theta 
\sin\phi \partial_\phi \,, \qquad \xi_3 =\partial_\phi \,, 
\end{align}
satisfying $[\xi_i,\xi_j ]=\epsilon_{ijk}\xi_k$. 

We now have two independent constants which specify the 3-dimensional base space. 
In appearance, this would give two different bases. However, we 
can always achieve $c_1=c_2=0$ by a rotation of the $\text{S}^2$, implying that 
the base space is unique up to isometry. A similar situation occurs
for five-dimensional minimal gauged supergravity for which 
the only way to describe AdS$_5$ in the timelike canonical form is the
fibration over the Bergmann space~\cite{Gutowski:2004ez}. 

The base space with $c_3=1$ corresponds to the one obtained by the $a\to 0$ limit 
of Kerr-Newman-AdS (to take this limit, the choice $c=-\ell/a$ in
(\ref{omega_y}) is convenient). 
In this case, the metric in the canonical form is given by 
\begin{align}
\D s^2 =- N\left(\D t+\frac{r^2 \sin^2\theta }{N\ell^2} \D y \right)^2 
+\frac{1}{N} \left[\D z^2+\frac{r^2}{\ell^2} U^2 \sin^2\theta (\D x^2+\D y^2 )
 \right]\,, 
\end{align}
where 
\begin{align}
t=\ti t \,, \qquad y=\ell \phi +\ti t \,, \qquad 
z=(r-M)\cos\theta \,, \qquad x=\ell [\alpha(r )+\beta (\theta )]\,,
\end{align}
with 
\begin{align}
 N=\left(1-\frac{M}{r}\right)^2+\frac{r^2}{\ell^2}\cos^2\theta \,,
 \qquad 
\alpha'(r) =\frac{r-M}{r^2U^2} \,, \qquad \beta '(\theta )=\cot\theta
 \,. 
\end{align}
This solution exemplifies that the static BPS metric is 
rotating in the canonical form.

\subsection{Second scaling limit: The C-metric}

The PD solution contains the C-metric as another subclass. 
After the inversion $q\to -1/q$, we perform the rescaling  
$(q,p,\sigma ,\tau )\to l ^{-1}(q, p, \sigma ,\tau )$, accompanied by
%----------------  Scaling   ----------------%
\begin{align}
n\to ln\,, \quad \varepsilon \to l^2 \varepsilon \,, \quad 
m\to l^3 m \,, \quad \mathsf Q+i\mathsf P  \to l^2 (\mathsf Q+i\mathsf P
 )
\,, \quad \alpha  \to \alpha + l^4 \mathsf P^2 \,.
\label{scaling2-param}
\end{align}
The $l\to \infty $ limit removes the rotation parameter and gives the C-metric
%----------------  C metric   ------------------%   
\begin{align}
 \D s^2 =\frac{1}{(p+q)^2} \left[
-Q(q) \D \tau ^2 +P(p)\D \sigma^2 +\frac{\D p^2}{P(p)} +\frac{\D q^2}{Q(q)}
\right] \,, \qquad 
\ma A =\mathsf Q q \D \tau +\mathsf Pp \D \sigma\,,
\label{Cmetric}
\end{align}
where
\begin{align}
P(p)&= (\alpha  -\Lambda/6)+2 n p-\varepsilon p^2+2m p^3 -(\mathsf
 Q^2+\mathsf P^2 )p^4\,, \\
Q(q)&= (-\alpha -\Lambda/6) +2n q +\varepsilon q^2 +2 m q^3 +(\mathsf
 Q^2+\mathsf P^2 ) q^4\,. 
\end{align}
The solution \eqref{Cmetric} with cosmological constant appeared for the first time in \cite{Plebanski:1976gy}.
For $\Lambda<0$ (the AdS C-metric, which is the case considered here), it describes either a pair
of accelerated black holes (with the acceleration provided by the pressure exerted by a strut), or a
single accelerated black hole, depending on the value of the acceleration parameter. A detailed
discussion of the physics described by the AdS C-metric can be found in \cite{Dias:2002mi}.

We now wish to obtain the conditions under which the solution \eqref{Cmetric} admits Killing spinors.
The integrability conditions for \eqref{Kill-spin} yield
%----------------  1st integrability   ------------------%
\begin{align}
\hat \nabla_{[\mu }\hat \nabla_{\nu ]} \epsilon =&
\left[\frac 18 C_{\mu \nu \rho\sigma}\Gamma^{\rho\sigma}+\frac{i}4 
\left(\nabla_\rho \ma F_{\mu \nu } +i \Gamma_5 \nabla_\rho\star 
 \ma F_{\mu \nu } \right)\Gamma^\rho 
\right.\nonumber \\& \left.~
-\frac{ i}{2\ell} (\ma F_{\mu \nu }+{i}\Gamma_5 \star
\ma F_{\mu\nu } +\Gamma_{[\mu }{}^\rho\ma F_{\nu ]\rho })\right.\nonumber \\
& \left. +\frac{1}{4}\left(E_{\rho
 [\mu }-\frac{1}{6}{E^\sigma }_\sigma g_{\rho [\mu }\right){\Gamma^\rho
 }_{\nu ]}-\frac{3i}{4} \left(
\nabla_{[\mu }\ma F_{\nu\rho]}+i\Gamma_5 \nabla_{[\rho }\star \ma F_{\mu\nu ] }
 \right)\Gamma^\rho  \right] \epsilon \,,
\label{minimalN2_KS_int}
\end{align}
where 
$\star \ma F_{\mu\nu }=(1/2)\epsilon_{\mu\nu\rho\sigma}\ma F^{\rho\sigma }$, and 
%-------------------  E(a,b)  ------------------%
\begin{align}
E_{\mu\nu }\equiv R_{\mu\nu }-2
 \left(\ma F_{\mu\rho }\ma F_\nu{}^\rho -\frac 14 g_{\mu\nu }\ma F_{\rho\sigma
 }\ma F^{\rho\sigma } \right)+\frac{3}{\ell^2} g_{\mu\nu } \,. 
\end{align}
When the bosonic equations of motion are satisfied, 
the last line of (\ref{minimalN2_KS_int}) drops out.
For the C-metric (\ref{Cmetric}), ${\rm det}([\hat \nabla_\mu,\hat \nabla_\nu ])=0$
is equivalent to ${\rm det}\Pi=0$, where
\begin{align}
 \Pi \equiv (\mathsf Q-i \Gamma_5 \mathsf
 P)(\ell^{-1}-\sqrt{Q}\Gamma_1-\sqrt{P}\Gamma_2 )
+\frac{i}{12} (P''+Q'')\Gamma_0\Gamma_1 \,.
\end{align}
Here we have employed the frame
\begin{align}
 e^0=\frac{\sqrt Q \D \tau}{p+q}  \,, \qquad 
 e^1=\frac{\D q}{(p+q)\sqrt Q} \,, \qquad 
e^2= \frac{\D p}{(p+q)\sqrt P}\,, \qquad 
e^3 = \frac{\sqrt P \D \sigma }{p+q} \,. 
\end{align}
The condition ${\rm det}\Pi=0$ boils down to
%--------------- BPS condition   ---------------%
\begin{align}
m[m^2 -(\mathsf Q^2+\mathsf P^2 )\varepsilon ]+2n (\mathsf Q^2+\mathsf
 P^2)^2& =0 \,, 
\label{BPS-Cmetric1}
\\
m^2 \left[\frac{\mathsf Q^2 -\mathsf P^2}{2\ell^2} + (\mathsf Q^2 +\mathsf P^2)
 \alpha  \right]
+ n^2 (\mathsf Q^2 +\mathsf P^2)^2&=0\,.
\label{BPS-Cmetric2}
\end{align}
In this case, it is straightforward to check that with
\begin{subequations}
\begin{align}
 f&=\frac{\mathsf Q\mathsf P}{\mathsf Q^2+\mathsf P^2}[m-(p-q)(\mathsf
 Q^2+\mathsf P^2)] \,, \\
g&=\frac{(\mathsf Q^2+\mathsf P^2)\varepsilon -m^2 +2m (\mathsf
 P^2q-\mathsf Q^2p)+2
(\mathsf Q^2+\mathsf P^2)(p^2\mathsf Q^2+q^2 \mathsf P^2)}{2 (\mathsf
 Q^2+\mathsf P^2)(p+q)} \,, \\
V&=\mathsf P \partial_\tau -\mathsf Q \partial_\sigma \,, \\
z&=\ell \frac{m^2+(\mathsf Q^2+\mathsf P^2)[m(p-q)+2pq(\mathsf
 Q^2+\mathsf P^2)-\varepsilon ]}{2 (\mathsf Q^2+\mathsf P^2) (p+q)} \,,
\end{align}
\end{subequations}
all the algebraic and differential bilinear equations (2.13)-(2.18) and (2.24)-(2.28)
in \cite{Caldarelli:2003pb} are satisfied. The conditions \eqref{BPS-Cmetric1},
\eqref{BPS-Cmetric2} are thus not only necessary, but also sufficient for supersymmetry of the
AdS C-metric\footnote{One might be concerned with equation \eqref{Delta-phi}, which actually
does not result from algebraic or differential constraints on the bilinears, but from
the additional condition (3.25) of \cite{Caldarelli:2003pb}.
However, this equation is satisfied since
one can verify that it is equivalent to the trace part of Einstein's
equations, provided \eqref{Delta-F}, \eqref{phi'} and \eqref{omega}
hold.}.

The canonical form \eqref{metric-BPS} of the AdS C-metric in the BPS limit is
given by
\begin{eqnarray}
\D s^2 &=& -\frac{\mathsf P^2Q(q)-\mathsf Q^2P(p)}{(p+q)^2}\left[\D t + \frac{\mathsf P^2Q(q)+\mathsf Q^2
P(p)}{\mathsf P^2Q(q)-\mathsf Q^2P(p)}\D y\right]^2 \nonumber \\
&+& \frac{(p+q)^2}{\mathsf P^2Q(q)-\mathsf Q^2P(p)}\left[\D z^2 + \frac{4\mathsf P^2\mathsf Q^2Q(q)P(p)}
{(p+q)^4}(\D x^2 + \D y^2)\right]\,,
\end{eqnarray}
where
\begin{displaymath}
t = \frac{\tau}{2\mathsf P}-\frac{\sigma}{2\mathsf Q}\,, \qquad y = \frac{\tau}{2\mathsf P}+
\frac{\sigma}{2\mathsf Q}\,, \qquad x = \alpha(q)+\beta(p)\,,
\end{displaymath}
with
\begin{eqnarray*}
\alpha'(q) &=& \frac{m\ell q + \ell(\mathsf P^2+\mathsf Q^2)q^2 - \frac{\ell m^2}{2(\mathsf P^2
+\mathsf Q^2)} + \frac{\ell\varepsilon}2}{2\mathsf P\mathsf QQ(q)}\,, \\
\beta'(p) &=& \frac{m\ell p - \ell(\mathsf P^2+\mathsf Q^2)p^2 + \frac{\ell m^2}{2(\mathsf P^2 +
\mathsf Q^2)} - \frac{\ell\varepsilon}2}{2\mathsf P\mathsf QP(p)}\,.
\end{eqnarray*}
In particular, we see that
\begin{equation}
e^{2\phi} = \frac{4\mathsf P^2\mathsf Q^2Q(q)P(p)}{(p+q)^4}\,,
\end{equation}
which is very similar to \eqref{sep-ansatz-phi}, but the product of two quartic functions
is now dressed with a factor $(p+q)^{-4}$.

\section{Supersymmetry of the general PD solution}
\label{sec:generalPD}

After having studied the two different scaling limits which remove either the
acceleration or the rotation parameter, we come now to the general PD solution
\eqref{metr-PD}, \eqref{F-PD}, with the aim to work out the necessary and sufficient
constraints imposed by the existence of Killing spinors. It turns out that the first integrability
condition ${\rm det}([\hat \nabla_\mu , \hat \nabla_\nu ])=0$ reduces again to 
the single equation ${\rm det}\Pi=0$.
The exact form of $\Pi$ is not illuminating so we do not display it
here and only show the final result. We find that ${\rm det}\Pi=0$
is equivalent to the two conditions
\begin{align}
&n [m^2+n^2-(\mathsf P^2+\mathsf Q^2) \varepsilon ]+2 m (\mathsf P^2+
 \mathsf Q^2) (\mathsf P^2-\alpha ) \nonumber \\
& \qquad +\frac{1}{\ell^2} \left[2n \mathsf P \mathsf Q +m (\mathsf P^2-
 \mathsf Q^2)\right]=0 \,, \label{BPS-gen-PD1}
\\
&(\mathsf P^2+\mathsf Q^2)[
m^2\mathsf P^2-n^2\mathsf Q^2- (m^2+n^2)\alpha 
]\nonumber \\ 
&\qquad+\frac{1}{\ell^2} \left[2mn\mathsf P\mathsf  Q+\frac{1}{2} (\mathsf
 P^2-\mathsf Q^2) (m^2-n^2)\right]
=0\,. \label{BPS-gen-PD2}
\end{align}
These equations constrain the parameters $\alpha$ and $\varepsilon$
to be functions of $m,n, \mathsf P, \mathsf Q$.

In order to recover the integrability conditions for the
two limiting cases, we have to be careful since  eqn.~(\ref{BPS-gen-PD1})
does not survive in these limits.
For the Carter-Pleba\'nski metric, we use the following relation instead
of (\ref{BPS-gen-PD1}), 
\begin{align}
[(m^2+n^2)-(\mathsf P^2+\mathsf Q^2) \varepsilon ]^2 -(\mathsf P^2+\mathsf
 Q^2)^2 \left[\frac{1}{\ell^4}+\frac{2}{\ell^2} (\mathsf P^2-\mathsf Q^2)
+4 (\mathsf P^2-\alpha )(\mathsf Q^2+\alpha )
\right] =0 \,, 
\nonumber 
\end{align}
which is obtained from (\ref{BPS-gen-PD1}) and (\ref{BPS-gen-PD2}). 
Then the limit (\ref{scaling1-param}) gives precisely eqn.~(\ref{integr-cond-PD}).
For the C-metric, we need to use the equation which eliminates $\alpha $ from (\ref{BPS-gen-PD1})
by using (\ref{BPS-gen-PD2}). Then the limit (\ref{scaling2-param})
recovers  (\ref{BPS-Cmetric1}) and (\ref{BPS-Cmetric2}). 

Provided the eqns.~\eqref{BPS-gen-PD1} and \eqref{BPS-gen-PD2} hold, a long but
straightforward calculation shows that with
\begin{subequations}
\begin{align}
f&=\frac{(\mathsf P^2+\mathsf Q^2)[c_- pq(p \mathsf Q+\mathsf Pq )-c_+(p \mathsf P-q\mathsf Q)]
-c_+c_- (p^2+q^2)}{(p^2+q^2)(\mathsf P^2+\mathsf Q^2)}
\,,  \\
g&= -\frac{c_-^2 pq+c_+^2}{(1-pq)(\mathsf P^2+\mathsf Q^2)}
\nonumber \\ & +\frac{
c_+ [\mathsf P (p^3+q)+\mathsf Q (p+q^3 )]
-c_-[\mathsf Pq^2 (p^3+q)-\mathsf Q p^2(p+q^3 )]
}{(1-p q )(p^2+q^2)}\,,  \\
 V&=c_+\partial_\tau -c_- \partial_\sigma \,,   \\ 
z&=\ell \frac{m^2+n^2 -(\mathsf P^2+\mathsf Q^2) (mp+nq ) }{1-pq}\,, 
\end{align}
\end{subequations}
where
\begin{align}
 c_+=m \mathsf P+n\mathsf Q\,, \qquad c_-=m\mathsf Q-n\mathsf P\,,
\end{align}
all the bilinear equations in \cite{Caldarelli:2003pb} are satisfied.
Eqs.~\eqref{BPS-gen-PD1} and \eqref{BPS-gen-PD2} are thus necessary and sufficient
for supersymmetry of the general PD solution.

\section{Euclidean case}
\label{sec:Euclidean}

Euclidean supersymmetric solutions and gravitational instantons are of importance 
due to their relevance for non-perturbative effects
in quantum gravity~\cite{Dunajski:2010zp,Gutowski:2010zs,Dunajski:2010uv}.
Moreover, the boundaries of BPS geometries that asymptote to Euclidean AdS
admit conformal Killing
spinors \cite{Klare:2012gn}, and provide thus possible backgrounds on which
Euclidean superconformal field theories can be defined. Such theories have been
attracting much attention recently in the context of localization
techniques \cite{Kapustin:2009kz,Jafferis:2010un,Hama:2010av,Hama:2011ea}.

In order to analytically continue the general PD solution \eqref{metr-PD}, \eqref{F-PD} to the Euclidean
case, we first proceed as in \cite{Griffiths:2005qp}, and explicitly include a parameter $\omega$
which represents the twist of the repeated principal null congruences. This is done by rescaling
\begin{displaymath}
p \to \omega^{1/2}p\,, \qquad q \to \omega^{-1/2}q\,, \qquad \tau \to \omega^{1/2}\tau\,, \qquad
\sigma \to \omega^{1/2}\sigma\,,
\end{displaymath}
\begin{displaymath}
m+in \to \omega^{-3/2}(m+in)\,, \qquad {\mathsf Q}+i{\mathsf P}\to\omega^{-1}({\mathsf Q}+i{\mathsf P})\,,
\qquad \varepsilon \to \omega^{-1}\varepsilon\,, \qquad k\to k\,,
\end{displaymath}
where the parameter $k$ is defined by $k=-\Lambda/6-{\mathsf P}^2+\alpha$. The resulting solution
can then be Wick-rotated by taking
\eq
\tau \to i\tau\,, \qquad \omega\to i\omega\,, \qquad n\to in\,, \qquad {\mathsf Q}\to i{\mathsf Q}\,,
\feq
which leads to
\begin{eqnarray}
\D s^2 = &&
\frac1{(1-pq)^2}\left\{\frac{Q(q)}{q^2-\omega^2p^2}(\D\tau -\omega p^2\D\sigma)^2 
+ \frac{q^2-\omega^2p^2}{Q(q)}\D q^2
\right.\nonumber \\ &&
+ \left.\frac{q^2-\omega^2p^2}{P(p)}\D p^2 
+ \frac{P(p)}{q^2-\omega^2p^2}(-\omega\D\tau + q^2\D\sigma)^2\right\}\,, 
\label{metr-PD-eucl}
\end{eqnarray}
\eq
{\cal F} = \frac{\mathsf{Q}(q^2+\omega^2p^2)-2\mathsf{P}pq\omega}{(q^2-\omega^2p^2)^2}\D q\wedge
(\D\tau - \omega p^2\D\sigma)
+ \frac{2\mathsf{Q}pq\omega-\mathsf{P}(q^2+\omega^2p^2)}{(q^2-\omega^2p^2)^2}\D p\wedge
(q^2\D\sigma-\omega\D\tau)\,, \label{F-PD-eucl}
\feq
where the structure functions are given by
\begin{eqnarray}
P(p) &=& k + 2\omega^{-1}np - \varepsilon p^2 + 2mp^3 + (\omega^2 k - \mathsf{P}^2 +
                  \mathsf{Q}^2 + \omega^2\Lambda/3)p^4\,, \nonumber \\
Q(q) &=& (-\omega^2 k + \mathsf{P}^2 - \mathsf{Q}^2) - 2mq + \varepsilon q^2 - 2\omega^{-1}nq^3
                  - (k + \Lambda/3)q^4\,.
\end{eqnarray}
Taking the tetrad frame
\begin{align}
e^1& =\sqrt{\frac{Q}{q^2-\omega^2 p^2}}
\frac{(\D \tau -\omega p^2 \D \sigma )}{1-pq}
\,, \qquad e^2 = \sqrt{\frac{q^2-\omega^2 p^2 }{Q}} \frac{\D q}{1-pq}
\,, \nonumber \\
e^3&= \sqrt{\frac{q^2-\omega^2 p^2}{P}}\frac{\D p}{1-pq} \,, \qquad 
e^4= \sqrt{\frac{P}{q^2-\omega^2 p^2}}\frac{(-\omega \D \tau +q^2 \D
 \sigma )}{1-p q} \,, 
\end{align}
one can define a self-dual two-form 
$\Omega=e^1\wedge e^2 + e^3\wedge e^4$ and $J^\mu{}_\nu :=g^{\mu\rho }\Omega _{\nu\rho }$.
It can then be shown that $J\cdot J=-1$
and the Nijenhuis tensor for $J$ vanishes, i.e., the almost complex
structure $J$ is integrable\footnote{This argument does not ensure the global existence of the complex
structure on the whole space. For instance, the manifold with $m=n=|\mathsf P|-|\mathsf Q|=0$ and
$\Lambda>0$ describes $\text{S}^4$, in which a global complex structure
cannot be defined.
We thank Yukinori Yasui for pointing this out. 
}
(see \cite{Mason:2010zzc} for a discussion of the Carter-Pleba\'nski family).
It should be noted that the two-form $\Omega $
fails to be closed so that it does not correspond to
a K\"ahler structure.

Taking into account the above rescaling and subsequent analytic continuation, the BPS conditions
\eqref{BPS-gen-PD1}, \eqref{BPS-gen-PD2} become
\begin{align}
&n [m^2-n^2-(\mathsf P^2-\mathsf Q^2) \varepsilon ]+2\omega m (\mathsf Q^2-
 \mathsf P^2) (\Lambda/6+k) \nonumber \\
& \qquad +\frac{\Lambda\omega}3 \left[2n \mathsf P \mathsf Q -m (\mathsf P^2+
 \mathsf Q^2)\right]=0 \,, \label{BPS-eucl1}
\\
&(\mathsf P^2-\mathsf Q^2)[
m^2\mathsf P^2-n^2\mathsf Q^2+ (m^2-n^2)(\omega^2 k + \Lambda\omega^2/6 - \mathsf P^2) 
]\nonumber \\ 
&\qquad+\frac{\Lambda\omega^2}3 \left[-2mn\mathsf P\mathsf  Q+\frac{1}{2} (\mathsf
 P^2+\mathsf Q^2) (m^2+n^2)\right]
=0\,. \label{BPS-eucl2}
\end{align}
It would be interesting to explicitly check from first principles, using the results of \cite{Dunajski:2010uv},
that the Euclidean solution \eqref{metr-PD-eucl}, \eqref{F-PD-eucl}, with the parameters given by
\eqref{BPS-eucl1}, \eqref{BPS-eucl2}, is indeed supersymmetric in
Euclidean gauged supergravity.

One can easily verify that the Euclidean PD solution is (anti-)self-dual if
\begin{align}
 m= \pm n \,, \qquad \mathsf Q=\pm \mathsf P \,,
\label{EucPD_SD}
\end{align}
for which 
$C_{\mu\nu\rho\sigma }=\pm (1/2) \epsilon_{\mu\nu \lambda\tau }C^{\lambda\tau}{}_{\rho\sigma}$
and 
$\ma F_{\mu\nu }=\pm(1/2)\epsilon_{\mu\nu\rho\sigma }\ma F^{\rho\sigma }$
are satisfied. When the $\text{(anti-)}$self-duality condition (\ref{EucPD_SD})
holds, the stress-energy tensor of the Maxwell field vanishes identically 
(hence the metric is Einstein), and
the BPS equations (\ref{BPS-eucl1}) and (\ref{BPS-eucl2}) follow automatically. 

Notice that the gravitational instantons \eqref{metr-PD-eucl}, \eqref{F-PD-eucl}, which have
$\text{U}(1)\times\text{U}(1)$ symmetry, generalize the ones with
$\text{SU}(2)\times\text{U}(1)$ symmetry constructed recently by Martelli, Passias and Sparks
in \cite{Martelli:2012sz}. It should be straightforward to recover the latter by a reasoning similar to that in
section 3 of \cite{Griffiths:2005qp}. We shall not attempt to do this here.

Note finally that in the subcase of the C-metric, a Euclidean continuation can be done trivially by
taking $\tau\to i\tau$, $\mathsf Q\to i\mathsf Q$.

\section{Final remarks}
\label{final}

Our main result in this paper are the necessary and sufficient conditions for
supersymmetry of the general Pleba\'nski-Demia\'nski solution \eqref{metr-PD}, \eqref{F-PD}.
We also considered two different scaling limits of this geometry, that lead to the
Carter-Pleba\'nski solution or the C-metric. For the former, the first integrability conditions
for Killing spinors were worked out in \cite{AlonsoAlberca:2000cs}, and we showed that
these are also sufficient for supersymmetry. The results obtained in our paper resolve thus
also some issues that remained open in the literature.

For these classes of solutions, we also revealed the general structure of the three-dimension\-al
base space over which a BPS geometry in ${\cal N}=2$ gauged supergravity is fibered,
and showed that for Reissner-Nordstr\"om-AdS, this base space is
unique up to isometry, in spite of the existence of two linearly independent Killing spinors.

Generically, the BPS solutions considered here are written in a rotating frame in the
canonical form \eqref{metric-BPS}, even when they are static.
This feature appears also in AdS$_5$ \cite{Gauntlett:2003fk}, and it would be clearly
desirable to understand this better.

The analytical continuation of our results to Euclidean signature yields gravitational instantons
that are supersymmetric in Euclidean gauged supergravity. The conformal boundaries of these
backgrounds provide new three-dimensional geometries on which Euclidean supersymmetric field
theories can be defined, since they admit conformal Killing spinors \cite{Klare:2012gn}.

Finally, our work may open the possibility to systematically construct generalizations of the
PD metric in matter-coupled gauged supergravity, where nontrivial scalar fields with a potential
are turned on, by using the recipe of \cite{Cacciatori:2008ek}.
Important examples of such solutions that have been constructed so far include
the rotating or NUT-charged black holes of \cite{Klemm:2011xw,Colleoni:2012jq},
the dilatonic C-metric without potential \cite{Dowker:1993bt}, the
PD solution conformally coupled
to a scalar in presence of a cosmological constant and a $\phi^4$ potential \cite{Charmousis:2009cm,Anabalon:2009qt}, as well as the solutions of \cite{Anabalon:2012ta}.
For obvious reasons, geometries of this type (in particular with a Liouville potential for the
dilaton) may be instrumental for the construction of black rings \cite{Emparan:2001wn} in
AdS$_5$ \cite{Caldarelli:2008pz}. We shall come back to these points in a forthcoming publication.

\acknowledgments

MN is thankful to Hideo Kodama, Yukinori Yasui and the participants of the
workshop
``Strong Gravity Beyond GR: From Theory to Observations'' (Lisbon)
for valuable comments.
MN also wishes to thank the Dipartimento di Fisica ``Galileo Galilei'' at
Padova University, the INFN Milano, and the Physics Department
at Milano University for kind hospitality during his stay.
This work was partially supported by INFN, MIUR-PRIN contract
2009-KHZKRX and by the MEXT Grant-in-Aid for Scientific Research on
Innovative Areas No. 21111006.

\end{document}